\documentclass[twocolumn,showpacs,floats,floatfix,superscriptaddress,aps,pra]{revtex4-1}
\usepackage{amsfonts}
\usepackage{amssymb}
\usepackage{amsmath}
\usepackage{calc}
\usepackage{graphicx}
\usepackage{bm}

\usepackage[normalem]{ulem}
\usepackage{amsmath,amssymb}
\usepackage{multirow}
\usepackage{xcolor,soul}
\usepackage{srcltx}
\usepackage{hyperref,graphicx}

\def\be{ \begin{equation}}
\def\ee{ \end{equation}}
\def\bea{ \begin{eqnarray}}
\def\eea{ \end{eqnarray}}
\def\bse{ \begin{subequations}}
\def\ese{ \end{subequations}}
\def\bc{ \begin{center}}
\def\ec{ \end{center}}

\begin{document}

\author{S. Longhi} 
\affiliation{Dipartimento di Fisica, Politecnico di Milano and Istituto di Fotonica e Nanotecnologie del Consiglio Nazionale delle Ricerche, Piazza L. da Vinci 32, I-20133 Milano, Italy}
\author{S. A. R. Horsley}
\affiliation{Department of Physics and Astronomy, University of Exeter, Stocker Road, Exeter, EX4 4QL}
\author{G. Della Valle} 
\affiliation{Dipartimento di Fisica, Politecnico di Milano and Istituto di Fotonica e Nanotecnologie del Consiglio Nazionale delle Ricerche, Piazza L. da Vinci 32, I-20133 Milano, Italy}

\title{Scattering of accelerated wave--packets}
  \normalsize


%
\bigskip
\begin{abstract}
\noindent  
Wave--packet scattering from a stationary potential is significantly modified when the wave--packet is subject to an external time-dependent force during the interaction. In the semiclassical limit, wave--packet motion is simply described by Newtonian equations and the external force can, for example, cancel the potential force making a potential barrier transparent.  Here we consider wave--packet scattering from reflectionless potentials, where in general the potential becomes reflective when probed by an accelerated wave--packet. In the particular case of the recently-introduced class of complex Kramers-Kronig potentials we show that a broad class of time dependent forces can be applied without inducing any scattering, while there is a  breakdown of the reflectionless property when there is a broadband distribution of initial particle momentum, involving both positive and negative components.
\end{abstract}



\maketitle

\section{Introduction}
Wave and particle scattering by a potential barrier or well is ubiquitous in many areas of classical and quantum physics \cite{R-2,R0,R0tris,R-1}. The phenomenology of scattering processes is greatly enriched when dealing with non-Hermitian \cite{ah1,ah2} or time-varying potentials, for example when the particle is exposed to time-dependent external fields while interacting with a static potential \cite{R0tris,R0quatris,R0bis}. Time-dependent scattering problems appear in several areas of physics, ranging from atomic and molecular physics \cite{R1,R2,R3,R3bis,R4} to condensed matter and mesoscopic systems \cite{R5,R6,R7,R8,R9,R10}. 
Dynamical scattering is also of major interest in connection with fundamental aspects of quantum mechanics, such as the problem of tunneling times \cite{Butt1,Butt2,Butt3,Butt4,Butt5} and models of classical and quantum chaos \cite{chaos2,chaos3,chaos4,book}.\\ 
 The majority of studies on dynamical scattering have been focused on time-periodic potentials, where Floquet theory applies and the scattering process is fully described by reflection and transmission amplitudes for elastic and inelastic scattering channels \cite{R0quatris,R0bis,R11,R12,Flo}. Floquet scattering is at the heart of some important physical effects, such as photon-assisted tunneling \cite{R5,r3,r4},
quantum pumps \cite{r5c,r5b,r5,r5graphene},  chaos-assisted tunneling \cite{book,r7,r8,r9,r11}, coherent destruction of
tunneling \cite{R0bis,r12}, quantum interference \cite{r13}, Floquet-Fano resonances \cite{R8,Flo}, field-induced barrier transparency \cite{Vor,lon}, etc. \\
Scattering from arbitrary time-periodic potentials or from stationary potentials with external non-periodic driving fields has received less attention so far \cite{R13,R14,R15,R16,R17,R18,R19,R20,R21}. The main reason is that the lack of time periodicity makes the scattering dynamics more involved, and in very few special cases an analytical treatment is available \cite{R14,R17,R18}. 
In time-dependent scattering theory, the scattering problem should be rather generally described in terms of localized wave--packets rather than extended plane waves \cite{R-1}. At initial time, i.e. ``in the distant past'', the wave--packet is well localized outside the interaction region, i.e. the region where the potential is non vanishing. After the scattering process, i.e. ``in the distant future'', one generally assumes that the wave--packet no longer interacts with the potential nor with the external driving forces, which are switched off after some time. The subsequent wave--packet evolution can be then used to determine probabilities for wave--packet reflection and transmission across the potential. However, even for wave--packet scattering from stationary potentials without driving forces, subtleties can arise when the initial wave--packet has a broad distribution in momentum space containing both positive and negative components \cite{Muga1}, i.e. when the incident wave--packet is not entirely right or left moving.\\

In this article we consider wave--packets scattered from a localized static potential barrier or well, that are accelerated by an external time-varying force while interacting with the potential. While in the semiclassical limit scattering is simply described by Newtonian equations of motion and the external force can be tailored to control wave--packet motion -- for example it can be used to effectively cancel the potential force making a potential barrier transparent -- more interesting physical results are found in the full wave regime, where the semiclassical limit does not provide an adequate description of the scattering problem and wave interference effects come into play. In particular, we consider scattering of accelerated wave--packets from potentials that are reflectionless, i.e. that do not reflect waves in the stationary (i.e. without external force) limit. There are several examples of non-reflecting potentials, such as the P\"oschl-Teller potential \cite{R22,R22bis,R22a}, the Kay--Moses potentials \cite{R23}, the complex absorbing potentials \cite{ah1,referee2,referee3,referee4}, and the Kramers-Kronig potentials \cite{R24}. Complex absorbing potentials have been introduced in numerical methods of reactive scattering and other molecular collisions to calculate continuum quantities  with finite grid or finite basis methods \cite{ah1,referee2,referee3,referee4,referee5}, avoiding or minimizing reflection effects at the boundaries. Kramers-Kronig potentials are a rather broad class of unidirectionally or bidirectionally reflectionless complex potentials, introduced by Horsley and coworkers in a recent work \cite{R24}, in which  the real and imaginary parts of the potentials are related one another by spatial Kramers-Kronig relations. Such potentials show rather interesting properties, such as unidirectional or bidirectional transparency, invisibility, perfect absorption, and robustness to spatio-temporal deformations, which  have been investigated in several recent works \cite{R21,R24,R25,R26,R27a,R28,R29,R30,R31,R32,R33,R34}. The main result of our study is that all classes of reflectionless potentials mentioned above  become reflective when probed by an accelerated wave--packet. In particular, for the class of Kramers-Kronig potentials breakdown of the reflectionless property is a more subtle effect and arises from broadband distribution of particle momentum involving both positive and negative components.

\section{Scattering of accelerated wave--packets from a static potential: model and basic equations}
We consider wave scattering from a one-dimensional localized potential $V(x)$ in the presence of a spatially-homogeneous time-dependent force $F(t)$. In dimensionless units ($\hbar=1$ and $m=1/2$), the Schr\"odinger equation for the wave--packet amplitude $\psi(x,t)$ reads
\begin{equation}
i \frac{\partial \psi}{\partial t}= -\frac{\partial^2 \psi}{\partial x^2} + V(x) \psi-F(t) x \psi.\label{eq:hamiltonian}
\end{equation}
The scattering potential $V(x)$ is generally assumed to be complex, leading to a non-Hermitian dynamics. Scattering from complex potentials is found in different physical contexts. Important examples include scattering from parity-time symmetric potentials \cite{Bender}, from complex absorbing potentials \cite{ah1,referee2,referee3,referee4}, and from Kramer-Kronig potentials \cite{R24}. The scattering problem can be formulated in two different reference frames. Besides the `laboratory' reference frame $(x,t)$,  scattering can be studied in the accelerated reference frame $x^{\prime}=x-x_0(t)$, $t^{\prime}=t$ where $x_0(t)$ is the classical trajectory of the particle due to the external force solely, i.e. $\ddot x_0(t)=(1/m) F(t)=2F(t)$ \cite{R3,chaos2,Vor,R21,Mois,uffa}. We then apply the Kramers-Henneberger transformation \cite{Kramers} to eliminate the final
term from equation (1): in the accelerated reference frame the Schr\"odinger equation (\ref{eq:hamiltonian}) is transformed to the one of a quantum particle without the external force (i.e. $F=0$) but with a time-dependent scattering potential $V^{\prime}(x^{\prime},t^{\prime})$ that drifts or oscillates in time according to $V^{\prime}(x^{\prime},t^{\prime})= V(x^{\prime}+x_0(t^{\prime}))$. While there exist generalizations of the Kramers-Henneberger transformation for spatially-inhomogeneous applied forces, such as those arising from non dipole approximation in light-atom interactions \cite{refereecazzo}, we do not consider here spatial inhomogeneities. Also, here we study the scattering process in the `laboratory' reference frame $(x,t)$.  \\
The scattering potential is assumed to be localized at around $x=0$ and to vanish sufficiently fast as $|x| \rightarrow \infty$ (short-rangle potential) so that scattering states are asymptotically plane waves. For slowly-decaying (long-range) potentials, like Kramers-Kronig potentials, one can envelope the potential by a sufficiently broad amplitude with rapidly decaying tails. In practice, one can assume $V(x)=0$ for $|x|>L$, where $L$ is a sufficiently large truncation length, so that without the external force ($F=0$) the scattering states of the Hamiltonian are plane waves with definite momentum for $|x|>L$.\\
For an arbitrary time-periodic force, the scattering problem should be described in terms of localized wave--packets \cite{note1}. Here we focus our attention to left incidence side, however a similar analysis holds for a wave--packet indecent from the right side of the scattering potential. At initial time $t=0$, the wave--packet $\psi(x,0)$ is  thus assumed to be fully localized outside the interaction region, i.e. $\psi(x,0) \simeq 0$ for $x>-L$.
The external force is switched on at $t=0$ and the wave--packet is accelerated, while it interacts with the static potential $V(x)$. After some time $t=T$, the force is switched off and the wave--packet dynamics is subsequently observed for long times.\par
To study the scattering process, let us first assume that there is not any scattering potential, i.e. $V(x)=0$. In this case, it is known that 
plane-wave solutions to Eq.(\ref{eq:hamiltonian}), which account for the effect of the external force $F(t)$, are given by the Gordon-Volkov states \cite{referee1}, which read explicitly
\begin{equation}
\psi_p(x,t)=\exp \left[ i\mathcal{P}(p,t)x-i \int_0^t d \xi \mathcal{P}^2(p,\xi) \right]\label{eq:eigenstates}
\end{equation}
where 
\begin{equation}
\mathcal{P}(p,t)=p+\int_0^t d \xi F( \xi)
\end{equation}
and $p$ is the wave particle momentum (wave number) at initial time $t=0$. Note that the effect of the external force is to change the particle momentum, from the initial value $p$ to the value $\mathcal{P}(p,t)$, according to the classical  law $d \mathcal{P}/dt=F(t)$. The most general solution to Eq.(\ref{eq:hamiltonian}) is given by a superposition of Gordon-Volkov states and reads
 \begin{eqnarray}
 \psi(x,t) & = & \int_{-\infty}^{\infty} dp G(p) \psi_p(x,t) \\
 & = & \int_{-\infty}^{\infty} dp G(p) \exp \left[ i\mathcal{P}(p,t)x-i \int_0^t d \xi \mathcal{P}^2 (p, \xi ) \right] \nonumber\label{eq:time_dependent_psi}
 \end{eqnarray}
where $G(p)$ is the momentum distribution of the initial wave--packet $\psi(x,0)$, i.e. $G(p)=(1/ 2 \pi) \int_{-\infty}^{\infty} dx \psi(x,0) \exp(-ipx)$. At times $t \geq T$, i.e. after switching off the external force, the wave--packet evolves according to
 \begin{eqnarray}
 \psi(x,t) & = & \int_{-\infty}^{\infty} dp G(p) \exp \left[ i(p+\Delta p) x-ip^2t - \right. \nonumber \\
   &- &  \left. i \Delta p (2p+\Delta p) (t-T) -  ip \phi_1 -i \phi_0 \right]
 \end{eqnarray}
 where we have set
 \begin{eqnarray}
 \Delta p & = & \int_0^T dt F(t) \\
  \phi_1 & = & 2 \int_0^T dt \int_0^t d \xi F( \xi) \\
  \phi_0 & = & \int_0^T dt \left( \int_0^t d \xi F( \xi) \right)^2\label{eq:phases}
 \end{eqnarray}
 Note that, provided that the impulse of the force over the interval $(0,T)$ vanishes, i.e. if the condition
 \begin{equation}
 \int_0^T dt F(t)=0 
 \end{equation}
 is met, at times $t \geq T$ one has
 \begin{equation}
 \psi(x,t)=\psi^{(free)}(x-\phi_1,t) \exp(-i \phi_0)
 \end{equation}
 where $\psi^{(free)}(x,t)$ describes the free-particle wave--packet evolution, i.e. uniform motion and quantum spreading of the wave--packet when $V=F=0$ in Eq.(\ref{eq:hamiltonian}). Therefore, for a vanishing impulse the effect of the external force is just to shift the wave--packet position as compared to the force-free motion, as one would expect from a simple semiclassical analysis. Moreover, provided that the additional condition
 \begin{equation}
 \int_0^T dt \int_0^t d \xi F(\xi)=0 
 \end{equation}
 is satisfied, one has $\psi(x,t)=\psi^{(free)}(x,t) \exp(-i \phi_0)$, i.e. apart from the inessential phase shift $\phi_0$ the external force does not change at all the evolution of the wave--packet as compared to the force-free dynamics.\par
 To study the scattering of a wave--packet, we decompose the wave amplitude $\psi(x,t)$ as a superposition of the scattering-free Gordon-Volkov states $\psi_p(x,t)$ by letting
 \begin{align}
 \psi(x,t)  &=  \int_{-\infty}^{\infty} dp \; c(p,t) \psi_p(t) \\
  &=\int_{-\infty}^{\infty} dp\; c(p,t) \exp \left[ i\mathcal{P}(p,t)x-i \int_0^t d \xi \mathcal{P}^2(p, \xi ) \right] \nonumber
 \end{align} 
   where $c(p,0)=G(p)$ is the momentum distribution of the incident wave--packet at initial time $t=0$. The evolution equations for the spectral amplitudes $c(p,t)$ are readily obtained after substitution of the Ansatz (12) into Eq.(\ref{eq:hamiltonian}) and read
   \begin{equation}
   i \frac{\partial c(p,t)}{\partial t}=\int_{-\infty}^{\infty} dq \; c(q,t) \tilde{V}(p-q) \exp [i \varphi(p,q,t)]\label{eq:c_evolution}
   \end{equation}
   where $\tilde{V}(q) \equiv (1/ 2 \pi) \int_{-\infty}^{\infty} dx V(x) \exp(-iqx)$ is the Fourier spectrum of the scattering potential $V(x)$ and where we have set
   \begin{align}
   \varphi(p,q,t)   &\equiv  \int_0^t d \xi [\mathcal{P}^2(p,\xi)-\mathcal{P}^2(q,\xi)] \;\;\;\;\;\;\;\; \\
    &=  \int_0^t d \xi \bigg\{ \left[ p+\int_0^{\xi} d \rho F( \rho)  \right]^2\nonumber\\
    &\hspace{3.5cm}- \left[ q+\int_0^{\xi} d \rho F( \rho)  \right]^2 \bigg\}.  \nonumber
   \end{align}
   Equation (\ref{eq:c_evolution}) is an integro-differential equation that governs the evolution of the spectral amplitudes $c(p,t)$ of force-driven plane waves $\psi_p(x,t)$ in the presence of the scattering potential.  It is remarkable that---despite the presence of the external time dependent forcing---the plane wave eigenstates (\ref{eq:eigenstates}) are coupled to one another by the Fourier amplitude $\tilde{V}(p-q)$, which depends on the value of the momenta at time $t=0$, i.e. before the force was applied.  Only the phase of the coupling between the different plane waves is affected by $F(t)$, and this is because the external force displaces all momentum states by the same amount so that $\mathcal{P}(p,t)-\mathcal{P}(q,t)=p-q$.     
   \par
   While the scattering equations (\ref{eq:c_evolution}) represent an exact result and holds for non-Hermitian (complex) potentials as well, unfortunately it is not amenable for an analytical study and only in special cases can it lead to exact results concerning the scattering (reflection) properties of the potential.
   
  \section{Semiclassical and fast periodic driving limits}
The problem of wave--packet scattering is greatly simplified in two limiting and well-established cases, which are briefly reviewed in this section. The first one is the semiclassical limit, whereas the second one is the high-frequency periodic driving case.

\subsection{Semiclassical limit}
The semiclassical limit of Eq.(\ref{eq:hamiltonian}) provides the simplest case where scattering of a wave--packet can be handled in a straightforward way. Although a semiclassical  description of wave--packet scattering is possible in the most general case of non-Hermitian Hamiltonians, i.e. for a complex scattering potential \cite{R35,R36,R37,R38,R39,R40}, its usefulness to describe wave--packet dynamics turns out to be quite limited for complex potentials since an infinite hierarchy of coupled equations for mean values of operators involving position $x$ and momentum $p_x=-i \partial_x$ is generally needed. Therefore we will limit here to consider the ordinary semiclassical limit assuming a real potential. In this case the exact equations for the temporal evolution of the mean values of wave--packet position $\langle x \rangle $ and momentum $\langle p_x \rangle $ are given  by the Ehernfest equations with $m=1/2$, which read
\begin{eqnarray}
\frac{d \langle x \rangle}{dt} & = & 2 \langle p_x \rangle \\
\frac{d \langle p_x \rangle}{dt} & = & F(t)- \left\langle \frac{\partial V}{\partial x} \right\rangle.
\end{eqnarray}
For a slowly-varying potential, the semiclassical limit is introduced as usual by assuming $\langle  ( \partial V / \partial x) \rangle \simeq  (\partial V / \partial x) ( \langle x \rangle)$, so that the mean position $\langle x \rangle$ satisfies the classical Newtonian equation of motion (with $m=1/2$)
\begin{equation}
\frac{1}{2} \frac{d^2 \langle x \rangle}{dt^2}=F(t)- \frac{\partial V}{\partial x} \left( \langle x \rangle \right).\label{eq:newton}
\end{equation}
In such a simple limiting case, the external force merely adds to the potential force to determine the trajectory of the mean wave--packet position according to the Newtonian equation (\ref{eq:newton}). For example, for a given initial wave--packet distribution, the external force $F(t)$ can be tailored to cancel the potential force, so as to effectively make a potential barrier transparent. Indicating by $x_0$ and $p_0$  the mean values of position and momentum of the wave--packet at initial time $t=0$, the external force must vary in time according to 
  \begin{equation}
  F(t)= \frac{\partial V}{\partial x} (x_0+2 p_0 t).\label{eq:newton2}
  \end{equation}
  As an example, Fig.1 shows the force-induced transparency of a Gaussian-shaped potential barrier as obtained from direct numerical simulations of the Schr\"odinger equation (\ref{eq:hamiltonian}) assuming an initial Gaussian wave--packet with an initial energy below the barrier. While in the absence of the external force the wave--packet is almost fully reflected from the potential barrier [Fig.1(a)], an external force tailored according to Eq.(\ref{eq:newton2}) enables complete crossing of the barrier [Fig.1(b)]. While in the semiclassical limit the barrier is made exactly transparent, quantum mechanically such a result is only an approximate one because the external force can cancel the scattering potential only locally. This can be seen by comparing the wave--packet probability distribution after barrier crossing  with the one corresponding to the free-particle motion (i.e. with $F=V=0$): clearly, a slight deviation between the two probability density distributions can be seen [compare solid and dashed curves in the middle panel of Fig.1(b)]. 
 \begin{figure}[htbp]
 \includegraphics[width=87mm]{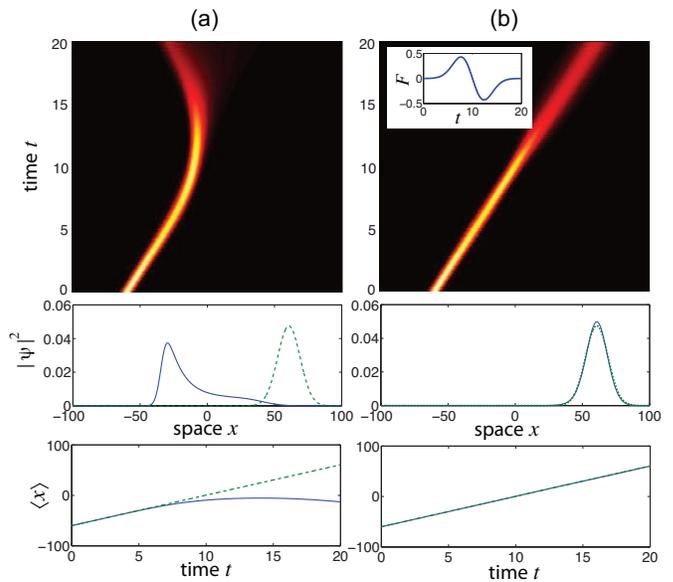}\\
  \caption{(color online) Force-induced barrier transparency based on the semiclassical analysis. (a) Numerically-computed scattering of a wave--packet from a Gaussian potential barrier $V(x)=V_0 \exp(-\alpha^2 x^2)$ in the absence of the external force ($F=0$) for parameter values $V_0=10$ and $\alpha=0.05$. The initial wave--packet distribution is $\psi(x,0)=\mathcal{N} \exp [-(x+d)^2/w^2+ip_0x]$ where $\mathcal{N}$ is the normalization constant and $d=60$, $w=5$, and $p_0=9$. The upper panel shows the wave--packet evolution (snapshots of $|\psi(x,t)|^2$ on a pseudo color map), the middle panel shows the detailed behavior of the probability density distribution $|\psi(x,t)|^2$ at final time $t=20$, whereas the bottom plot shows the trajectory of the wave--packet center of mass $\langle x (t) \rangle$. The dashed curves in the middle and bottom panels correspond to the free-propagation wave--packet ($V=0$). (b) Same as (a), but for the external force $F=F(t)$ tailored according to Eq.(\ref{eq:newton2}). The behavior of the force $F(t)$ is depicted in the inset of the upper panel. The dashed lines in the middle and bottom panels, corresponding to the free-particle regime ($F=V=0$), are almost overlapped with the solid curves.}
\end{figure}

\subsection{High-frequency periodic forcing}
Another special case, where a simple analytical treatment is available, corresponds to a periodic and high-frequency external force. In this regime Floquet formalism can be applied and the scattering problem is usually studied in the Kramers-Henneberger reference frame $x^{\prime}=x-x_0(t)$, $t^{\prime}=t$ \cite{R3,chaos2,Vor,Mois}. In the high-frequency limit, the rapidly oscillating potential $V(x^{\prime},t^{\prime})=V(x^{\prime}+x_0(t^{\prime}))$ can be replaced at leading order by its time average over one oscillation cycle, i.e. the scattering problem basically reduces to the ordinary scattering of a static effective potential given by 
\begin{equation}
V^{(av)}(x)=\frac{1}{\tau} \int_0^{\tau} dt V(x+x_0(t)),\label{eq:V_av}
\end{equation}
 where $\tau=2 \pi / \omega$ is the oscillation period of the force \cite{chaos2,Vor,Mois}. Here we show that the same result can be obtained in the laboratory reference frame $(x,t)$ using the general scattering equations (\ref{eq:c_evolution}). To this aim, let us note that Eq.(\ref{eq:c_evolution}) can be cast in the equivalent form
 \begin{equation}
   i \frac{\partial c(p,t)}{\partial t}=\int_{-\infty}^{\infty} dq \; c(q,t) \hat{W}(p-q,t) \exp \left[ i(p^2-q^2) t \right]\label{c_limit}
   \end{equation}
 where we have set 
 \begin{equation}
 \hat{W}(q,t) \equiv \tilde{V}(q) \exp \left[ q x_0(t) \right]
 \end{equation}
 and 
 \begin{equation}
 x_0(t) \equiv \frac{1}{m} \int_0^t d \xi \int_0^{\xi} d \rho F ( \rho)=2 \int_0^t d \xi \int_0^{\xi} d \rho F ( \rho).
 \end{equation}
 Clearly, $\hat{W}$ is the Fourier spectrum of the oscillating potential $V(x+x_0(t))$, i.e. $V(x+x_0(t))=\int dq \tilde{V}(q) \exp[iq(x+x_0(t))]=\int dq \hat{W}(q,t) \exp(iqx)$. In the high-frequency limit, i.e. for a rapidly oscillating force, the amplitude $c(p,t)$ is not able to follow the rapid changes of $\hat{W}(q,t)$ over one oscillation cycle, and therefore at leading order one can replace, in Eq. (\ref{c_limit}), $\hat{W}(q,t)$ with its time-average over one oscillation cycle (rotating-wave approximation). After averaging, one then obtains
   \begin{equation}
   i \frac{\partial c(p,t)}{\partial t} \simeq \int_{-\infty}^{\infty} dq \; c(q,t) \tilde{V}^{(av)}(p-q,) \exp \left[ i(p^2-q^2) t \right]
   \end{equation}
where $\tilde{V}^{(av)}(q)$ is the Fourier spectrum of the cycled-averaged potential $V^{(av)}(x)= (1 / \tau) \int_0^{\tau} dt V(x+x_0(t))$. The above integro-differential equation is precisely the equation that one would obtain when considering the scattering problem from a stationary potential $V^{(av)}(x)$ using standard plane-wave expansion method. Thus, a rapidly-oscillating periodic force is equivalent to an effective reshaping of the scattering potential, which is at the basis of important effects in atomic physics such as adiabatic stabilization of atoms in intense high-frequency laser fields \cite{R3}, dynamical tunneling \cite{chaos2} and field-induced barrier transparency \cite{Vor}. 

\begin{figure}[htbp]
 \includegraphics[width=87mm]{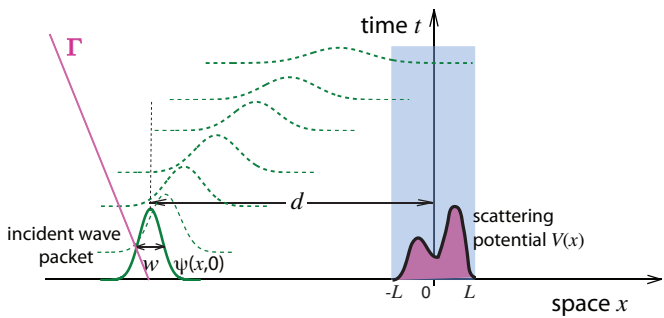}\\
  \caption{(color online) Schematic of wave--packet scattering (left incidence side) from a static potential. The temporal behavior of the wave function $\psi(x,t)$ is monitored  along the straight line $\Gamma$ of space-time plane $x=-d+v_dt$, where $v_d$ is a drift velocity. For a reflectionless potential the condition (\ref{eq:packet_decay}) given in the text should be fulfilled for any negative drift velocity $v_d<0$ and rather arbitrary initial wave--packet shape.}
\end{figure}
 \begin{figure}[htbp]
 \includegraphics[width=87mm]{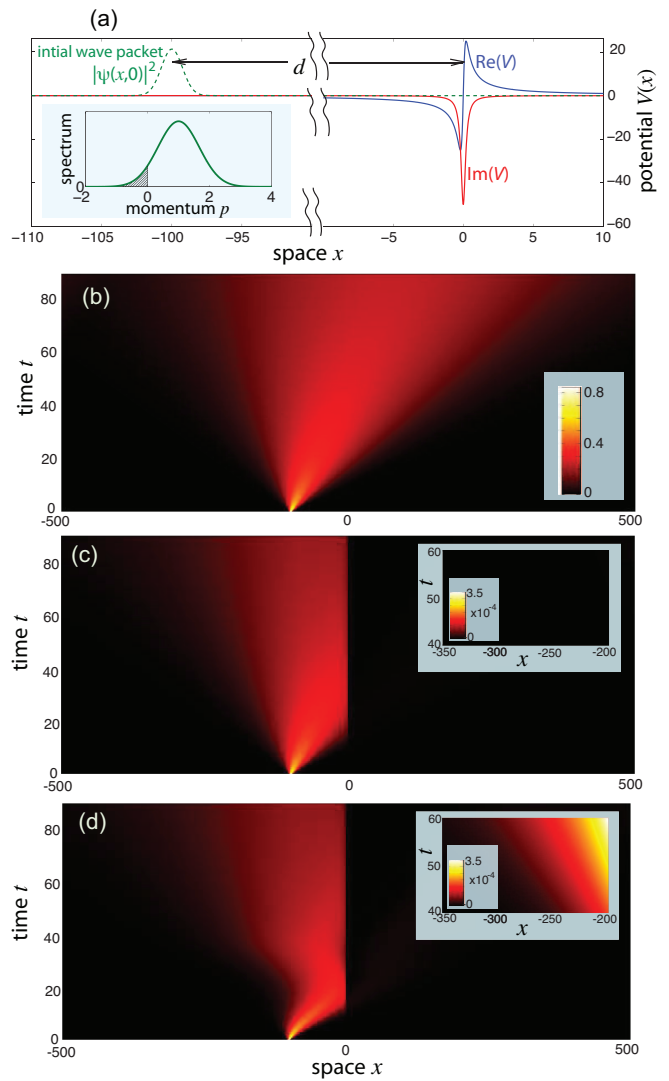}\\
  \caption{(color online) Numerical simulations showing scattering of a wave--packet with broad momentum distribution from a single-pole dissipative Kramers Kronig potential [Eq.(\ref{eq:V_example})] for parameter values $V_0=10$ and $\alpha=0.2$. (a) Behavior of the real and imaginary parts of $V(x)$ (solid curves). To make the potential short-range, $V(x)$ is enveloped by the broad super-Gaussian amplitude $\exp[-(x/b)^4]$ with $b=60$.  The initial wave--packet is $\psi(x,0) \propto \exp[-(x+d)^2/w^2+ip_0x]$ with $d=100$, $w=1.2$ and $p_0=1$. The dashed curve shows the initial wave--packet probability distribution $|\psi(x,0)|^2$, whereas the inset on the left bottom depicts the momentum distribution of the wave--packet. Note that negative momentum components are non-negligible. (b) Free evolution of the wave--packet $\psi^{(free)}(x,t)$ in the absence of the external force and of the scattering potential, i.e. for $F=V=0$ [snapshot of $\sqrt{|\psi(x,t)|}$ on a pseudo color map]. (c) Same as (b), but with the scatting potential $V(x)$ and without the external force ($F=0$). (d) Same as (b), but when the wave--packet is accelerated by the force given by Eq.(\ref{eq:F_example}) with parameter values $T=40$ and $F_0=0.25$. The insets in (c) and $(d)$ show the behavior of $|\psi^{(free)}(x,t)-\psi(x,t) \exp(i \phi_0)|$ in the upper-left region of space-time $(x,t)$, far from the scattering region and at times after switching off the external force. }
\end{figure}
\section{Scattering of an accelerated wave--packet from reflectionless potentials}
An important class of potentials that have been long known in quantum mechanics and optics (see e.g. \cite{R22,R22bis,R23}), are those special potential profiles that do not reflect waves at all. The reflectionless nature of such  potentials are a wave interference effect, so that the various scattering paths destructively interfere resulting in the absence of a reflected wave. Another class of potentials that do not reflect waves, or minimize wave reflection,  are so-called complex absorbing potentials (see \cite{ah1,referee2,referee3,referee4} and references therein), which are introduced in numerical methods of reactive scattering and other molecular collisions.
 More recently, a new class of reflectionless potentials, so-called Kramers-Kronig potentials, has been recently introduced in Ref.\cite{R24} and attracted great interest in the past few years \cite{R25,R26,R27a,R28,R29,R30,R31,R32,R33,R34}. Such potentials are complex, i.e. they correspond to a non-Hermitian Hamiltonian, and their profile is such that the real and imaginary parts of the potential are related to one another by a Hilbert transform. This is equivalent to saying that the Fourier spectrum $\tilde{V}(q)$ of the potential $V(x)$ has a one-sided support, for example $\tilde{V}(q)=0$ for $q<0$. The Kramers-Kronig potentials are generally undirectionally reflectionless, i.e. wave reflection is observed for one incidence side but not for the opposite one. The main question we wish to address in this section is the following one: does a reflectionless potential remain relfectionless when the scattering process is assisted by an external time-dependent force? A partial answer to this question has been recently given in Ref.\cite{R21}, where it was shown that Hermitian potentials like P\"oschl-Teller or Kay-Moses potentials loose their reflectionless property when they oscillate in time, i.e. when an external driving force is applied. Similarly, reflectionless complex absorbing potentials are expected to become reflective in time-periodic problems, such in photoionization problems of atoms or molecules in strong laser fields.
 This can be seen, for instance, by considering the periodic and high-frequency forcing regime discussed in Sec.III.B. In this limit the problem is basically reduced to wave scattering from the stationary cycled-average potential $V^{(av)}(x)$, defined by Eq.(\ref{eq:V_av}). Clearly, the strong reshaping of the potential introduced by time averaging destroys the special form of P\"oschl-Teller, Kay-Moses or complex absorbing complex potential profiles, thereby loosing their reflectionless property. On the other hand, Kramers-Kronig potentials show a kind of supremacy since, in the limit of high-frequency driving, if $V(x)$ is a Kramers-Kronig potential, $V^{(av)}(x)$ is also a Kramers-Kronig potential with one-sided Fourier spectrum. In Ref.\cite{R21} it was shown that the property of the Kramers-Kronig potentials to remain reflectionless under temporal deformations is a rather general feature, i.e. it holds beyond the limiting case of periodic and high-frequency forcing, when considering wave--packets with positive momentum components solely (for left incidence side). In view of such a robustness, complex Kramers-Kronig potential could be useful, for example, as artificial potentials to impose absorbing boundary conditions without spurious reflection in simulations of large-scale strongly coupled scattering problems encountered in molecular physics. However, strictly localized wave--packets have a broad momentum spectrum that can be non--zero for both positive and negative momentum, making the analysis of Ref.\cite{R21} not exhaustive. Even for scattering by static potentials, it is known that some subtleties arise when the wave--packet is not fully directed against the potential barrier, i.e., when the initial momentum distribution has negative components \cite{Muga1}. In addition, wave--packets with broad momentum distributions can show anomalous behaviors such as quantum back flows \cite{Butt5,back1,back2,back3,back4,back5,back6}. Therefore a more in depth study is required to reveal how an accelerated wave--packet with negative-momentum components is scattered from a Kramers-Kronig potential. The main result that we shall prove here is that, while in the absence of the accelerating force a Kramers-Kronig potential is reflectionless even for wave--packets that are not fully directed toward the potential, i.e. with non-negligible negative momentum components, reflection can be observed when wave--packets with non-negligible negative-momentum components are accelerated by the external force toward the potential. 

\subsection{Definition of reflectionless potential for accelerated wave--packets}
As a preliminary remark, let us provide a consistent definition of reflectionless potentials when the scattering problem is formulated in terms of wave--packets. Let us assume a scattering potential with limited support in the spatial region $(-L,L)$, or a short-range potential with interaction length $ \sim L$ \cite{note2}, and an initial wave--packet distribution $\psi(x,0)$ which is localized on the left side of the scattering potential, at a distance $d$ larger (or possibly much larger) than the interaction length $L$; see Fig.2 for a schematic. The localization length $w$ of the wave--packet is assumed to be much smaller than $d$, so that the wave--packet is entirely localized  far from the scattering region. The wave--packet spectrum $G(p)=(1/ 2 \pi) \int_{-\infty}^{\infty} dx \psi(x,0) \exp(-ipx)$ is centered at the positive mean momentum $p=p_0>0$, however we generally assume that negative momentum components are non-negligible. Wave--packet localization near $x=-d$, with localization length $w$ much smaller than $d$, is obtained rather generally by assuming $G(p)=G_0(p) \exp(ipd)$, with $G_0(p)$ peaked near $p=p_0$ and slowly varying with respect to $p$ on the scale of the order $ \sim 1/d$. In the absence of the scattering potential and without the external force, i.e. for $V=F=0$, the free evolution of the wave--packet is given by
\begin{equation}
\psi^{(free)}(x,t)=\int_{-\infty}^{\infty} dp G(p) \exp(ipx-ip^2t).
\end{equation}
Let us consider the behavior of the wave--packet along the straight line $\Gamma$ of space-time defined by the relation $x=-d+v_dt$, where $v_d$ is either a negative or positive drift velocity. For large times, using the stationary phase method one readily obtains the following asymptotic behavior of $\psi^{(free)}(x=-d+v_d t,t) \equiv \psi^{(free)}_{v_d}(t)$
\begin{equation}
\psi^{(free)}_{v_d}(t) = \sqrt{\frac{\pi}{t}} G_0 \left( \frac{v_d}{2}\right) \exp \left( it \frac{v_d^2}{4}-i \frac{\pi}{4} \right) +o \left( \frac{1}{\sqrt{t}}\right)
\end{equation}
which decays as $ \sim 1 / \sqrt{t}$ at large times provided that $G_0(v_d/2) \neq 0$. Note that, for $v_d<0$ and provided that the wave--packet spectrum $G(p)$ is composed by positive components solely, i.e. $G(p) \simeq 0$ for $p<0$, $\psi^{(free)}_{v_d}(t)$ decays faster in time than $ \sim 1 / \sqrt{t}$, a signature that the wave--packet is entirely composed by progressive waves (so-called `right-moving' wave--packet). Let us now consider the case where there is not the scattering potential ($V=0$), but the wave--packet is accelerated by the external force $F(t)$. We assume that conditions (9) and (11) for the force are satisfied, so that according to Eq.(10) one has $\psi(x,t)=\psi^{(free)}(x,t) \exp(-i \phi_0)$ at times $t \geq T$, where the phase $\phi_0$ is defined by Eq.(\ref{eq:phases}). In the presence of a scattering potential, it is therefore reasonable to say that $V(x)$ is a reflectionless potential (for left incidence side) whenever, for any negative drift velocity $v_d<0$ and rather arbitrary initial wave--packet shape, one has
\begin{equation}
\psi^{(free)}_{v_d} (t)-\psi_{v_d}(t) \exp(i \phi_0) \sim o \left( \frac{1}{\sqrt{t}} \right) \label{eq:packet_decay}
\end{equation}
as $t \rightarrow \infty$. 
 
 \subsection{Scattering of wave--packets from a Kramers-Kronig potential without external force}
 Let us first consider wave--packet scattering from a Kramers-Kronig potential $V(x)$ without the external force, i.e. for $F(t)=0$. In this case, the potential is reflectionless for an arbitrary initial wave--packet, that is even for wave--packets with non-vanishinig negative momentum components, as it should be for a reflectionless potential. To prove such a statement, let us indicate by $\varphi_1(x,p)$ and $\varphi_2(x,p)$ the two linearly-independent scattering states of the Hamiltonian $\hat{H}=-d^2/dx^2+V(x)$ with the same energy $E=p^2$ and with the asymptotic behavior \cite{note2}
 \begin{eqnarray}
 \varphi_1(x,p)= \left\{  
 \begin{array}{cc}
 \exp(ipx)+r_-(p) \exp(-ipx) & x<-L \\
 t(p) \exp(ipx) & x>L
 \end{array}
 \right.
 \end{eqnarray}
 \begin{eqnarray}
 \varphi_2(x,p)= \left\{  
 \begin{array}{cc}
 t(p) \exp(-ipx) & x<-L \\
 \exp(-ipx)+r_+(p) \exp(ipx) & x>L \;\;\;
 \end{array}
\right.
 \end{eqnarray} 
 where $t(p)$, $r_-(p)$ and $r_+(p)$ are the spectral transmission and reflection (for left $r_-$ and right $r_+$ incidence sides) amplitudes and $p>0$. For a Kramers-Kronig potential with $\tilde{V}(q)=0$ for $q<0$, one has $r_-(p)=0$ for any $p>0$. The initial wave--packet distribution can be written as a suitable superposition of the scattering states $\varphi_1(x,p)$ and $\varphi_2(x,p)$ with spectral amplitudes $G_1(p)$ and $G_2(p)$. The wave--packet evolution at successive times is then given by
 \begin{eqnarray}
 \psi(x,t) & = & \int_0^\infty dp \left[ G_1(p) \varphi_1(x,p)+G_2(p) \varphi_2(x,p) \right]  \nonumber \\
 & \times & \exp(-ip^2 t)
 \end{eqnarray}
where the spectral amplitudes $G_1(p)$ and $G_2(p)$ are determined by the spectrum $G(p)=G_0(p) \exp(ipd)$ of $\psi(x,0)$. Taking into account the asymptotic form Eqs.(27) and (28) of the scattering states, one can readily check that $G_1(p)$ and $G_2(p)$ should be chosen as follows
\begin{widetext}
\begin{equation}
G_1(p)  =  G_0(p) \exp(ipd)+ \frac{G_0(-p)r_+(p) \exp(-ipd)}{r_-(p)r_+(p) -t^2(p)} \; , \;\; 
G_2(p)  =  - \frac{t(p) G_0(-p) \exp(-ipd)}{r_-(p)r_+(p)-t^2(p)}
\end{equation}
\end{widetext}
($p>0$). Using the method of the stationary phase, at long times the value of the wave--packet amplitude $\psi_{v_d}(t)=\psi(x=-d+v_dt,t)$ on the line $\Gamma$ can be readily obtained from Eqs.(29) and (30). For $v_d<0$ one obtains
\begin{eqnarray}
\psi_{v_d}(t)  & = & \left[ G_0 \left( \frac{v_d}{2} \right) +r_- \left( -\frac{v_d}{2} \right) G_0 \left( -\frac{v_d}{2} \right)  \right] \nonumber \\
& \times & \sqrt{\frac{\pi}{t}} \exp \left(i t \frac{v_d^2}{4}-i \frac{\pi}{4} \right)+ o \left( \frac{1}{\sqrt{t}}\right).
\end{eqnarray}
According to the definition of reflectionless potentials for wave--packet scattering given in Sec.IV.A [see Eq.(\ref{eq:packet_decay})], a comparison of Eqs.(25) and (31) clearly shows that the stationary potential $V(x)$, without any external force, is reflectionless if and only if $r_-(p)=0$ for any $p>0$. Therefore, a Kramers-Kronig potential is reflectionelss for arbitrary wave--packets, i.e. even for those comprising negative momentum components.

 \subsection{Scattering of accelerated wave--packets from a Kramers-Kronig potential}
The main feature of Kramers-Kronig potentials, recently shown in Ref.\cite{R21}, is that they remain reflectionless under temporal deformations of the potential. However, to what extent and under which conditions the reflectionless property is conserved, was not fully investigated in such a previous work. Here we show that a sufficient condition for the potential to remain reflectionless is that the incident wave--packet should be composed by positive momentum components solely, i.e. the constraint $G(p)=0$ for $p \leq 0$ should be imposed. However, for highly-localized wave--packets with a broad momentum distribution comprising non-negligible negative components, reflection can be instead observed.\\
\subsubsection{Reflectionless property for positive-momentum wave--packets}

Let us assume that the initial wave--packet is entirely `right-moving', i.e. it comprises positive momentum components solely, $G(p)=0$ for $p \leq 0$, and that the Kramers-Kronig potential is reflectionless for left-incidence side, i.e. $\tilde{V}(q)=0$ for $q<0$. 
The exact solution to the scattering problem is governed by the integro-differential equation (\ref{eq:c_evolution}). Clearly, since $c(p,0)=G(p)=0$ for $p \leq 0$ and $\tilde{V}(q)=0$ for $q<0$, from Eq.(\ref{eq:c_evolution}) it readily follows that, at any time $t>0$, one has {\em exactly } $c(p,t)=0$ for $p \leq 0$, i.e. at any time $t$ the wave--packet is  a `right-moving' wave--packet. As stated previously, this is simply because the effect of the uniform force $F(t)$ is to displace every momentum eigenstate by the same amount over time, so that the difference in momentum between any two modes $\mathcal{P}(p,t)-\mathcal{P}(q,t)$ is time independent. Given the fact that $c(p,t)=0$ for $p \leq 0$, an application of the stationary phase method confirms that the wave amplitude $\psi_{v_d}(t)=\psi(x=-d+v_dt,t)$ on the line $\Gamma$ decays in time faster than  $ \sim 1/ \sqrt{t}$ for any $v_d<0$, like for free-space propagation, and thus the reflectionless condition (\ref{eq:packet_decay}) is surely met. This result is in agreement with previous analysis of Ref.\cite{R21}.\par
In general, if we write our Hamiltonian as $\hat{H}=\hat{H}_{0}+V(x)$ where $\hat{H}_{0}$ is a Hamiltonian that is time dependent over an interval $[0,T]$, and $V(x)$ is a Kramers--Kronig potential, then there is a formal condition on $\hat{H}_{0}$ so that $V$ remains reflectionless for positive momentum wave--packets.  We give this condition in appendix~\ref{ap:A}.  In appendix~\ref{ap:B} we give an example of another specific form of time dependent Hamiltonian that also has this property.

\subsubsection{Reflection of wave--packets with broad momentum distribution}

Let us assume that the initial wave--packet $\psi(x,0)$ is tightly localized around $x=-d$ with a broad momentum distribution $G(p)$ centered at a positive value $p=p_0>0$ but with non-negligible negative components [see the inset in Fig.3(a)]. In this case, the proof of reflectionless scattering given above is not valid anymore, and the potential is expected to lose its reflectionless property for accelerated wave--packets. We can gain some qualitative physical insights into the scattering process of tightly-localized wave--packets using the superposition principle. Namely, we write the incident wave--packet as the interference of two wave--packets, $\psi(x,0)=\psi^{(r)}(x,0)+\psi^{(l)}(x,0)$, where 
\begin{eqnarray}
\psi^{(r)}(x,0) &  \equiv  & \int_0^\infty dp G(p) \exp(ipx) \\
\psi^{(l)}(x,0)  & \equiv &  \int_{-\infty}^0 dp G(p) \exp(ipx)
\end{eqnarray}
 are `right-moving' and `left-moving' wave--packets. As shown above, the `right-moving' wave--packet does not give rise to reflection, therefore we may focus our attention to the evolution of the `left-moving' wave--packet, i.e. the wave--packet with solely negative momentum content. According to the definition of reflectionless potential for a wave--packet given in Sec.IV.A, it is clear that the potential is not reflectionless whenever it can modify the evolution of $\psi^{(l)}(x,t)$ as compared to its free evolution. In the absence of the external force, the `left-moving' wave--packet, being initially localized far apart from the scattering region and moving on the left side, does not interact with the potential, and therefore we retrieve the result of Sec.IV.B that a non-accelerated wave--packet is not reflected, even if it comprises negative momentum components. However, an external force $F>0$ shifts the momentum distribution of the wave--packet to positive values and correspondingly  $\psi^{(l)}(x,t)$ can be brought close (or even beyond) the interaction region $x=0$. Since the impulse of the force vanishes, in the time interval where $F<0$ the momentum distribution is shifted toward negative values, and the resulting wave--packet $\psi^{(l)}(x,t)$  is again a `left-moving'  wave--packet, but with a profile which has been modified by the interaction with the potential near $x=0$ in earlier times. Therefore we expect violation of condition (\ref{eq:packet_decay}). Such a simple physical picture also indicates that breakdown of the reflectionless property is expected provided that the semiclassical trajectory of the wave--packet $\psi^{(l})$, induced by the external force solely,  gets close or even crosses the interaction region $x=0$.\par
 To check the predictions of the theoretical analysis,  the scattering of accelerated wave--packets from a Kramers-Kronig potential has been simulated by direct  numerical integration of the Schr\"odinger equation (\ref{eq:hamiltonian}) using a standard pseudo-spectral split-step method. As a Kramers-Kronig potential, we used the purely dissipative single-pole potential
 \begin{equation}
 V(x)=\frac{V_0}{x+i \alpha}\label{eq:V_example}
 \end{equation}
($V_0,\alpha>0$), which is shown in Fig.3(a). In the simulations, the potential (\ref{eq:V_example}) is enveloped by a broad super-Gaussian profile that makes $V(x)$ a short-range potential. The external force $F(t)$ used in the numerical simulations, satisfying the conditions (9) and (11), is given by
\begin{equation}
F(t)= \left\{ 
\begin{array}{cc}
F_0 \cos ( 2 \pi t /T) & 0 < t <T \\
0 & t>T
\end{array}
\right.\label{eq:F_example}
\end{equation} 
with $T=40$ and varying amplitude $F_0$. The initial wave--packet is Gaussian shaped with a broad momentum distribution, with positive mean value $p_0=1$ and tightly  localized in space at a distance $d=100$ from the interaction region $x=0$ [Fig.3(a)].
\begin{figure}[htbp]
 \includegraphics[width=87mm]{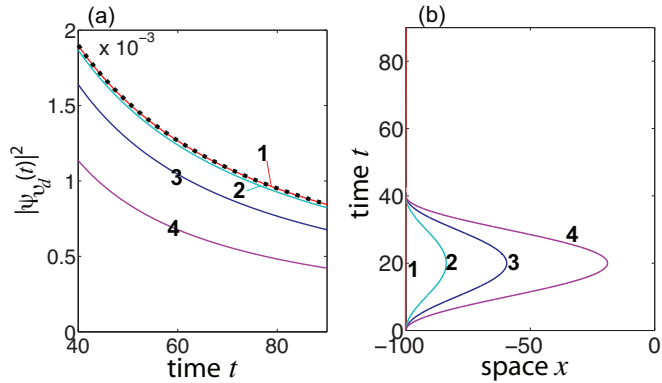}\\
  \caption{(color online) (a) Behavior of the probability density $|\psi_{v_d}(t)|^2$ along the line $\Gamma$: $x=-d+v_d t$ for $v_d=-0.2$, $d=100$ and for increasing values of the force amplitude $F_0$. Curve 1: $F_0=0$; curve 2: $F_0=0.1$; curve 3: $F_0=0.25$; curve 4: $F_0=0.5$. Squares, overlapped with curve 1, show to the behavior of $|\psi_{v_d}(t)|^2$ for the freely evolving wave--packet, i.e. for $F=V=0$. (b) Trajectories $X_0(t)=-d+(1/m) \int_0^t d \xi \int_{0}^{\xi} d \rho F(\rho)$ of a classical particle, initially at rest and at the position $x=-d$, under the action of the external force solely for increasing values of the force strength $F_0$. }
\end{figure}
Figures 3(b-d) show the numerically-computed evolution of the wave--packet in three distinct cases: the free evolving wave--packet, i.e. when $V=F=0$ [Fig.3(b)]; scattering of the non-accelerated wave--packet, i.e. when $F=0$ but $V \neq 0$ [Fig.3(c)]; and scattering of the accelerated wave--packet with $F_0=0.25$ [Fig.3(d)]. The insets in Figs.3(c) and (d) depict the space-time behavior of the difference $\Delta(x,t)=|\psi^{(free)}(x,t)-\psi(x,t) \exp(i \phi_0)|$ on the far left spatial region of the scattering potential and at times $t>T$, where $\psi^{(free)}(x,t)$ is the free wave--packet evolution of Fig.3(b). A vanishing value of $\Delta(x,t)$ corresponds to reflectionless wave--packet scattering, while non-vanishing values of $\Delta$ are the signature that condition (\ref{eq:packet_decay}) is violated and some kind of reflection has occurred. Clearly, in Fig.3(c) there is not wave--packet reflection, whereas reflection can be appreciated in Fig.3(d). The strength of reflection increases as the amplitude $F_0$ of the external force increases, as shown in Fig.4. Figure 4(a) shows the detailed numerically-computed evolution of the probability density $|\psi_{v_d}(t)|^2$ along the line $\Gamma$ $x=-d+v_dt$ for a drift velocity $v_d=-0.2$ and for increasing values of the force amplitude $F_0$. Figure 4(b) depicts the corresponding trajectory $X_0(t)=x_0(t)-d=(1/m)\int_0^t d\xi \int_0^{\xi} d \rho F(\rho)-d$ of a classical particle under the action of the external force solely, initially at rest and at position $x=-d$. While for $F_0=0$ the curve $|\psi_{v_d}(t)|^2$ is overlapped with the free wave--packet evolution curve, indicating the absence of reflection, as $F_0$ increases above zero deviations from the free wave--packet evolution dynamics is clearly observed. Note that the deviations are stronger for trajectories that gets closer to the scattering region $x=0$, according to the predictions of the theoretical analysis.
\subsubsection{Quantum-optical analogy}
Quantum scattering off a potential under the influence of a time-dependent spatially homogeneous force is analogous to scattering of monochromatic TE-polarized optical waves from an inhomogeneous dielectric medium (a slab) in the $(x,z)$ plane, where the quantum potential $V(x)$  is replaced by the refractive index distribution $n(x)$ of the medium, the particle momentum is analogous to the incidence angle, and the time axis $t$ in Fig.2 is analogous to the spatial coordinate $z$ (see, for example, \cite{R30}). The effect of the external force is emulated by considering wave scattering in the Kramers-Henneberger (non inertial) reference frame, where the potential becomes time-dependent, corresponding to a $z$-dependent refractive index distribution $n=n(x-x_0(z))$ \cite{R41,R42}. While tailoring complex  potentials and observing reflectionless properties of Kramers-Kronig potentials for quantum particles (such as cold atoms in an optical potential) could be challenging in experiments, scattering of optical waves from engineered dielectric media could provide an experimentally accessible testbed for the observation of the reflectionless property of Kramer-Kronig potentials and breakdown of such a property for wave packets with broad angular spectrum. Recent experiments in optics and microwaves are providing first evidences of reflectionless and invisibility properties of synthetic Kramers-Kronig potentials \cite{R33,R34,R43}.  An experiment aimed to observe the effect shown in Fig.3 could be envisaged as follows. A point-like light emitter, such as fluorescence from pumped quantum dots or active atoms, is placed close to the optical interface. The broad angular spectrum of the emitted wave effectively emulates quantum scattering of a wave packet with a broad momentum distribution, comprising negative-momentum components. While for a straight interface [no external force, Fig.3(b)] the light pattern on the left half plane, where the point-like source is placed, is not influenced by the Kramers-Kronig-shaped refractive index profile, a deformed interface in the neighbor of the emitter, corresponding to an external time-dependent force [Fig,3(d)], is expected to change the light pattern in the same half plane.

  \section{Summary and Conclusions}
  
We investigated the behavior of wave--packets propagating through a region of space where the potential $V(x,t)$ is the sum of a static part and time--varying part, the latter representing the effect of a uniform external force $F(t)$.  Our focus was on the question: how can the external force be applied to manipulate the propagation of a pulse through the static potential? 
\par
In the semiclassical limit (where the static potential varies slowly in space) the motion of the center of the packet can be described using Newton's equations, and the external force $F(t)$ can be chosen to locally cancel the gradient of the potential, making the potential transparent.  The wave-packet thus propagates as if in a region of empty space.  In contrast, when the external force $F(t)$ varies in a rapid periodic fashion, the wave-packet propagates as if in a modified (time averaged) static potential.  This averaging can (in principle) turn a highly scattering potential into a weakly scattering one.
\par
In the final section we considered the effect of the external forcing on a wave--packet propagating through a potential that would ordinarily be reflectionless.  For a real valued reflectionless potential such as those of the P\"oshl--Teller type, the external force will lead to some reflection of the pulse (a result which follows from the earlier findings of~\cite{R21}).  However, for the complex reflectionless potentials of the Kramers--Kronig type, this effect is somewhat more subtle, and rather surprising.  If the incident wave--packet is composed of only positive momenta, then the potential remains reflectionless, whatever the external forcing $F(t)$.
\par
A simple explanation of this finding is as follows: it is because the effect of the external uniform force is to rigidly translate the whole momentum distribution of the pulse over time, returning every mode to its initial momentum after time $T$.  Consequently the difference in momentum between any two given modes has the same value throughout the time the force is applied.  Because Kramers--Kronig potentials cannot convert positive momentum modes to negative momentum ones, they therefore cannot couple the final negative momentum states to the initial positive ones, whatever the time dependence of the force connecting them.  Appendices \ref{ap:A}--\ref{ap:B} demonstrate that this can be generalized to other time dependent potentials that also do not induce scattering.
\par
Meanwhile, narrow wave--packets (composed of both positive and negative momenta) behave differently.  These packets will still not scatter from a static Kramers--Kronig potential, despite their momentum content.  However, the action of an external time--dependent force can lead to an apparent scattering of these narrow packets.  This is not (as is typical) due to a conversion from positive to negative momentum (which is still ruled out), but is simply due to a change in the negative momentum content of the pulse.
\par
In our study we have not considered the case of inhomogeneous spatial forces which arises, for instance, in laser-atom interaction beyond the electric dipole approximation \cite{refereecazzo}. However, spatial dependence of the vector potential in the Hamiltonian is expected to spoil the reflectionless property of Kramers-Kronig potentials, regardless of the spectral broadness of the incident wave packet, simply because the form of space-time dependence of the force does not correspond to a rigid translation of the wave packet momentum distribution over time, as discussed in Appendices A and B.

%
%
\appendix
\section{A general argument for the absence of scattering in Kramers--Kronig potentials \label{ap:A}}
\par
The finding presented in the main manuscript, where, whatever the applied time dependent force $F(t)$, a positive momentum wave--packet does not scatter from a Kramers--Kronig potential can be generalized to give a pair of conditions on any applied time dependent potential.
\par
We write the full Hamiltonian as
\begin{equation}
	\hat{H}=\hat{H}_0+V(x)
\end{equation}
where $\hat{H}_0$ is a rather arbitrary and Hermitian Hamlitonian, which is time dependent during the interval $[0,T]$, and $V(x)$ is a Kramers--Kronig potential.  The solution to the Schr\"odinger equation is now transformed as $\psi=\hat{U}(t)\phi$ where $\hat{U}^\dagger\hat{U}=\boldsymbol{1}$ and ${\rm i}\hat{U}^{\dagger}\partial_{t}\hat{U}=\hat{H}_{0}$.  The Schr\"odinger equation for $\phi$ is then
\begin{equation}
	{\rm i}\frac{\partial\phi}{\partial t}=\hat{U}^{\dagger}V(x)\hat{U}\phi
\end{equation}
We assume that before and after the time dependence of $\hat{H}_{0}$ has been turned off, $\hat{H}_{0}$ is the Hamiltonian of free space, with plane wave eigenstates.  We therefore use the momentum representation of $\phi$
\begin{equation}
	\phi(x,t)=\int dp \; c(p,t)\exp({\rm i}px)\label{eq:momentum_rep}
\end{equation}
finding that the evolution of the coefficients $c(p,t)$ obey
\begin{multline}
	{\rm i}\partial_{t}c(p,t)=\\
	\int \frac{dq}{2 \pi} \left[\int\exp(-{\rm i}px)\hat{U}^{\dagger}V(x)\hat{U}\exp({\rm i}qx)dx\right] c(q,t)
\end{multline}
If the potential is to remain reflectionless then we must have
\begin{equation}
\int\exp(-{\rm i}px)\hat{U}^{\dagger}V(x)\hat{U}\exp({\rm i}qx)dx=0\;\;\;(p<0, q>0)\label{eq:cond1}
\end{equation}
for all times.  In addition the time evolution of $\hat{H}_0$ cannot have converted any positive momentum eigenstates to negative momentum ones, meaning that we also must have
\begin{equation}
\int\exp(-{\rm i}px)\hat{U}(T)\exp({\rm i}qx)dx=0\;\;\;(p<0, q>0)\label{eq:cond2}
\end{equation}
In general we can add a Kramers--Kronig potential to any such time dependent Hamiltonian $\hat{H}_{0}$ that satisfies condtions (\ref{eq:cond1}) and (\ref{eq:cond2}), and a wave--packet composed of positive momenta at $t=0$ will still only contain positive momenta after $\hat{H}_{0}$ returns to the free space Hamiltonian at $t=T$.
\par
Evidently the Hamiltonian of the main manuscript satisfies both the above conditions.   Condition (\ref{eq:cond1}) is satisfied because the transformation $\hat{U}$ acts to shift the momentum by a time dependent constant and adds a phase (cf. (\ref{eq:time_dependent_psi})), which means that (\ref{eq:cond1}) is proportional to $\tilde{V}(p-q)$.  Condition (\ref{eq:cond2}) is satisfied simply because the momentum eigenstates are shifted back to their original position in momentum space after time $T$.  As appendix~\ref{ap:B} demonstrates, there are other choices of time dependent Hamiltonian that also satisfy (\ref{eq:cond1}--\ref{eq:cond2}). 

\section{A simple example of another family of time dependent potentials that do not induce scattering\label{ap:B}}
Consider the following time dependent Hamiltonian
\begin{equation}
	\hat{H}=-\frac{\partial^{2}}{\partial x^{2}}-\frac{{\rm i}\alpha(t)}{2}\left(x\frac{\partial}{\partial x}+\frac{\partial}{\partial x}x\right)+V(x)\label{eq:ham_ex}
\end{equation}
where $\alpha(t)$ is an arbitrary real function, and $V(x)$ is a non--Hermitian potential.  Such a form for the Hamiltonian can be obtained e.g. from a transformation of the time-depenent simple harmonic oscillator Hamiltonian.  In the momentum representation (\ref{eq:momentum_rep}) the Schr\"odinger equation (\ref{eq:hamiltonian})  for the above Hamiltonian takes the form
\begin{multline}
	{\rm i}\left[\partial_{t}+\frac{\alpha(t)}{2}\left(q\partial_{q}+\partial_{q}q\right)\right]c(q,t)\\
	=c(q,t)q^{2}+\int \frac{dp}{2\pi} c(p,t)\tilde{V}(q-p)\label{eq:no_trans}
\end{multline}
This equation can be simplified after a change of variables from $q$ to $q'=\beta(t)q$ (where $\beta(t)=\exp(-\int^{t}\alpha(\xi)d\xi)$).  This transforms (\ref{eq:no_trans}) into
\begin{multline}
	{\rm i}\partial_{t}c(q',t)=c(q',t)\left[\left(\frac{q'}{\beta(t)}\right)^{2}-{\rm i}\frac{\alpha(t)}{2}\right]\\
	+\int \frac{dp'}{2\pi\beta(t)} c(p',t)\tilde{V}(\beta^{-1}(t)(q-p))
\end{multline}
In the case where $V(x)$ is a Kramers--Kronig potential, $\tilde{V}$ vanishes for negative values of the argument.  Given that $\beta$ is a positive number, this means that the evolution of the negative momenta are thus completely uncoupled from the positive momenta.  This has the consequence that a wave--packet composed of initially positive momenta will also not be scattered by the time dependence present in the Hamiltonian (\ref{eq:ham_ex}), whatever the form of $\alpha(t)$.  In this case the Hamiltonian $\hat{H}_{0}$ satisfies conditions (\ref{eq:cond1}--\ref{eq:cond2}) because it acts to scale the distribution of positive momenta, but never converts from positive to negative. This could have been anticipated from the classical equation of motion $\dot{p}=-\alpha p$ derived from the equivalent classical Hamiltonian $H_0=p^{2}+\alpha(t)xp$.
%



\begin{thebibliography}{31}
\bibitem{note1}
The ordinary formulation of the scattering problem in terms  of reflection/transmission amplitudes of various (elastic or inelastic) scattering channels for plane wave incidence, commonly used to study scattering processes from static potentials of for time-periodic potentials, does not have a straightforward generalization to an arbitrary time-dependent  potential. In this case the scattering problem is at best formulated in terms of scattering of localized wave--packets rather than plane waves.
\bibitem{note2}
A Kramers-Kronig potential shows an algebraic decay as $x \rightarrow \pm \infty$, i.e. it does not have a limited support and in some cases it is even not a short-range potential (see \cite{R25,R28}). However, in practice one can always truncate the potential, assuming $V(x)=0$ for $|x|>L$ and $L$ large enough \cite{R24}, or one can multiply $V(x)$ by an envelope with e.g. a broad super-Gaussian amplitude that makes the potential short-range.
\end{thebibliography}
\end{document}